\documentclass[12pt]{article}
\usepackage{epsfig}

\usepackage{relsize}

\def\beq{\begin{equation}}
\def\eeq#1{\label{#1}\end{equation}}
\def\eeqn{\end{equation}}

\def\beqa{\begin{eqnarray}}
\def\eeqa#1{\label{#1}\end{eqnarray}}
\def\eeqan{\end{eqnarray}}

\let\bar=\overbar

\def\Dslash{\not{\hbox{\kern-4pt $D$}}}
\def\dslash{\not{\hbox{\kern-2pt $\del$}}}

\def\msb{{\bar{\ssstyle M \kern -1pt S}}}

\def\Title#1{\begin{center} {\Large {\bf #1} } \end{center}}

\begin{document}

\Title{ $\bf{D^{0}-\bar{D}\,^{0}}$ Mixing: An Overview }

\bigskip\bigskip


\begin{raggedright}  

{\it J\"org Marks\index{Marks, J.} for the {\slshape B\kern-0.1em{\smaller A}\kern-0.1emB\kern-0.1em{\smaller A\kern-0.2em R}} Collaboration\\
Physikalisches Institut der Universit\"at Heidelberg \\
Philosophenweg 12,
D-69120 Heidelberg, GERMANY}
\bigskip\bigskip
\end{raggedright}

\begin{abstract}
Recently, the $B$ factory experiments {\slshape B\kern-0.1em{\smaller A}\kern-0.1emB\kern-0.1em{\smaller A\kern-0.2em R}} and Belle as well as the CDF collaboration found evidence for mixing in the $D$ meson system.
The current status (beginning of summer 2008) of the experimental results of $D^0$ mixing is summarized.
\end{abstract}
\section{Introduction}
The most 
surprising
 result of last year's spring conferences on particle physics was the report of 
the evidence for $D^0-\bar{D}\,^0$ mixing by both the 
{\slshape B\kern-0.1em{\smaller A}\kern-0.1emB\kern-0.1em{\smaller A\kern-0.2em R}}
and Belle collaborations~\cite{d0discoverBABAR,d0discoverBELLE}.
At the end of 2007, the CDF collaboration found evidence for $D^0$ mixing~\cite{d0discoverCDF} 
in a different environment but in the same decay channel as 
{\slshape B\kern-0.1em{\smaller A}\kern-0.1emB\kern-0.1em{\smaller A\kern-0.2em R}}.

The first mixing results were obtained in the neutral kaon system~\cite{kdiscover}, 50 years ago.
Mixing in the $B^0$ system~\cite{bddiscover} was established in 1987 and measurements of the 
mixing parameters in the $B_s$ system were
published in 2006 by the CDF and D0 collaborations~\cite{bsdiscover}.    

Within the Standard Model (SM), mainly the first two generations contribute to the $D^0$ mixing 
and the mixing parameters and CP violation are expected to be very small. 
The observation of the $D^0$ mixing completes the picture of quark 
mixing, since the $D^0$ system, in contrast to the other three neutral systems, involves down-type quarks
in the mixing loop.
On the other hand, non-Standard Model processes could enhance either the mixing or
the CP violation or both. Therefore, the measurement of large mixing parameters 
in the $D^0$ system or sizable CP violation would be a strong indication for New Physics (NP).

In this paper, we present an overview of $D^0$ mixing. 
After an introduction to the charm mixing phenomenology and 
analysis techniques, 
results of the mixing para\-meters and 
CP violation as related to mixing are summarized. They are obtained 
from hadronic two-body, multi-body final
states and from quantum correlated $D^0$ decays of
the experiments {\slshape B\kern-0.1em{\smaller A}\kern-0.1emB\kern-0.1em{\smaller A\kern-0.2em R}},
Belle, Cleo and CDF.  
Mixing results from semileptonic $D^0$ decays can be found elsewhere~\cite{semileptonicbabar,semileptonicbelle}.
\section{Mixing formalism and notation}
The neutral mesons are created as eigenstates of the strong interaction and can be distinguished by an 
internal quantum number, e.g. charm. Due to the weak interaction, an initially defined state of $| D^0 \rangle$ or 
$| \bar{D}\,^0 \rangle$ will evolve with time into a mixture of $D^0$ and $\bar{D}\,^0$.
The time evolution can be described by an effective weak Hamiltonian in the time dependent Schr\"odinger equation
\begin {displaymath}
i \frac{\partial}{\partial t} {D^0(t) \choose \bar{D}\,^0(t) }  = (M - \frac{i}{2}\Gamma) {D^0(t) \choose \bar{D}\,^0(t) }  \; ,
\end {displaymath}
where $M$ and $\Gamma$ are mass and decay width matrices.
The solutions of the Schr\"odinger equation are the mass eigenstates 
\begin {displaymath}
\vert D_{1,2} \rangle = p \vert D^{0} \rangle \mp q  \vert \bar{D}^{0} \rangle \; , 
\end {displaymath}
with
$p^2 + q^2 =1$.
They  are linear combinations of the flavor eigenstates  $D^{0}$ and $\bar{D}^{0}$. The mass eigenstates, $\vert D_1 \rangle $ 
and $\vert D_2 \rangle $,
propagate independently in time with their own lifetime $\Gamma_{1,2}$ and mass $M_{1,2}$ 
\begin {displaymath}
 \vert D_{1,2} (t) \rangle =  e ^{  -i(M_{1,2} -i \Gamma_{1,2} /2) t}  \vert  D_{1,2} (t=0) \rangle  \; \; . 
\end {displaymath}

The ratios $x \equiv  2 ( M_1-M_2) / (\Gamma_1 + \Gamma_2) $ and $ y \equiv (\Gamma_1 - \Gamma_2) /(\Gamma_1 + \Gamma_2) $
are related to the difference in lifetime and mass of the mass eigenstates.
These variables are referred to as mixing parameters and are the observables to be measured.

The probability $I$ to find the state $\vert D^0 \rangle$ from an initial state $\vert D^0 \rangle $ after a time $t$ is
\begin {displaymath}
I (D^0 \rightarrow D^0;t):= \; \vert \langle D^0  \vert   D^0(t) \rangle   \vert^2 =  \frac{e^{-\Gamma t}}{2} [\cosh(\Gamma t) + \cos (\Gamma t) ]
\end {displaymath}
and the one to find a  $\vert \bar{D}\,^0 \rangle$ is
\begin {displaymath}
I (D^0 \rightarrow \bar{D}\,^0;t):= \; \vert \langle \bar{D}\,^0  \vert   D^0(t) \rangle   \vert^2 =  \frac{e^{-\Gamma t}}{2}  \left\vert \frac{p}{q}\right\vert^2 [\cosh( \Gamma t)  - \cos ( \Gamma t) ]   \; \; .
\end {displaymath}
After a certain time, the opposite flavor component appears.
Mixing will occur if either the mass difference $x$ or the lifetime difference $y$ of the two  states is non-zero.
Depending on the size of $x$ and $y$, an oscillating behavior can be observed, e.g. as in the case of the $B_s$ system.

If CP violation is neglected
the state $\vert D_1 \rangle$  ($\vert D_2 \rangle$) is CP-even (CP-odd).
However,
CP violation in $D$ mixing can be parametrised in terms of the quantities 
$r_m  \equiv \left\vert \frac{ q}{p}\right\vert$ and $ \phi_f \equiv \arg (\frac{q \bar{A_f}}{p A_f})$, 
where  $A_f \equiv \langle f | \mathcal{H}_D | D^0 \rangle $ ( $\bar{A}_f \equiv \langle f | \mathcal{H}_D | \bar{D}\,^0 \rangle $) is the amplitude of a $D^0$ ($\bar{D}\,^0$) to decay into a final state $f$ and $\mathcal{H}_D$ is
the Hamiltonian of the decay. A value of $r_m \neq 1$ would indicate CP violation in mixing. A non-zero value of
$\phi_f$ would indicate CP violation in the interference between mixing and decay. 

There are two contributions to the charm mixing processes, the short range
box contributions and a long range part with on- and off-shell intermediate 
hadronic states. 
In contrast to the other neutral
systems, the box diagrams of the $D^0$ system involve loops of down type quarks. 
Due to the GIM mechanism and the CKM matrix suppression, the lowest-order 
short distance calculation gives tiny results ($x_{\rm{box}} = \mathcal{O} (10^{-5})$
and $y_{\rm{box}} = \mathcal{O} (10^{-7})$~\cite{theorie-box}). 
Theoretical predictions for the long-range contributions are very difficult, as the $c$ quark is either not heavy enough or  
too heavy to be treated by the different theoretical frameworks. 
Evaluations in the Operator Product Expansion (OPE) framework have shown that the mass and lifetime 
differences are enhanced with increasing order
in the OPE~\cite{theorie-0} compared to an OPE quark level analysis~\cite{theorie-NP0}. This
yields $ \mathcal{O} \approx 10^{-3}$ for both $x$ and $y$.

In the SM, CP violation in the charm sector is expected to be small and to be below 
the sensitivity of the experiments. Any measurement of CP violation in $D^0$ mixing would be 
a strong indication of NP.

Calculations within the SM  have large uncertainties
and, therefore, the ability to detect NP contributions is limited.
Nevertheless, the $D^0$ mixing measurements allow for a restriction of the parameter space of NP models.
Golowich et al. (summary in~\cite{theorie-golowich}) have explored which NP models yield sizable values for $x$ and $y$. 
They addressed 21 NP models from various areas and found that in case of 17 models tightened restrictions on the 
model parameter space can be placed~\cite{theorie-NP2}. 
\section{Experimental techniques}
The most significant measurement of the mixing parameters was performed by the
{\slshape B\kern-0.1em{\smaller A}\kern-0.1emB\kern-0.1em{\smaller A\kern-0.2em R}} collaboration
in the decay $D^0 \rightarrow K \pi$~\cite{d0discoverBABAR}. Therefore, basic ideas and techniques of mixing analyses
are explained using this decay channel.
\subsection{\label{flavortagging}Flavor tagging}
In order to perform a mixing measurement, the inital state has to be prepared and tagging of the flavor at production time is required.
The standard technique in charm physics is the use of $D^* \rightarrow D^0 \pi $ decays \footnote{The charge conjugate modes are included throughout this paper.}. Here, the charge of the pion $\pi_{\rm{tag}}$ 
determines the flavor of the $D^0$ at production time.

The flavor at decay time is determined by the final state particle properties. Lets consider the decay of an initial $D^0$ which
decays without mixing as $D^0 \rightarrow K^- \pi^+$. In case of mixing, the $D^0$ converts to a $\bar{D}\,^0$ and 
decays as $D^0 \rightarrow K^+ \pi^-$. Therefore, the charge of the $K$ determines the flavor at decay time and thus if mixing 
occurred.  Events of the decay mode
 $D^0 \rightarrow K^- \pi^+$ are classified as right-sign (RS) and events of the decay mode $D^0 \, (\rightarrow \bar{D}\,^0) \rightarrow K^+ \pi^-$ are classified
as wrong-sign (WS).

Beside the Cabbibo-favored (CF) processes mentioned above, the $D^0$ also decays doubly Cabbibo-suppressed (DCS) as $D^0 \rightarrow K^+ \pi^-$.
As a consequence, the WS event sample contains not only events which have undergone mixing followed by a CF decay, but also DCS decays. 
The rate of the DCS events relative to the CF events is suppressed by a factor $\tan^4 (\theta_C)$, where $\theta_C$ is the Cabbibo angle. The rate of events with mixing is about another factor hundred smaller. Moreover, both decay amplitudes interfere.
\subsection{\label{mixingparameter}Extracting mixing parameters} 
While the DCS events decay just exponentially,
the mixed decays should have a more complex time structure due to the mixing process.  
The time evolution of the WS decay rate $T_{\rm{WS}}$ can be approximated by
\begin{equation}
\label{timeevolution}
T_{\rm{WS}} (t) \propto  e^{-\Gamma t}  \left( \underbrace{R_{\rm{D}}}_I  + \underbrace { \sqrt{R_{\rm{D}}} \,  y' \, \Gamma t }_{II}  +  \underbrace { \frac{x'^2 + y'^2}{4}  (\Gamma t)^2 }_{III} \right) 
\end{equation}
where  $R_{\rm{D}}$ is the rate of the DCS events, $y'$ and $x'$ denote the mixing parameters and $\Gamma t$ is the time in units 
of the $D^0$ decay time. CP conservation and small mixing parameters are assumed. 
There are three contributions to $T_{\rm{WS}}$,  
the DCS decays (I), the interference of the DCS and mixed decays (II) 
and the mixed decays (III).
Each has a different time dependence. 
Measuring the time dependence of the WS decay rate allows to determine the mixing parameters $x$ and $y$.
Because of the strong phase difference between the CF decay amplitude and the DCS decay amplitude
the mixing parameters are only defined up to a phase factor $\delta_{K \pi}$. Therefore, the mixing parameters are measured as 
$x' = x \; \sin \delta_{K\pi} + y \; \cos \delta_{K\pi} $ and $y' = -x \; \sin \delta_{K\pi} + y \; \cos \delta_{K\pi}$.  
The phase $\delta_{K \pi}$ depends on the decay mode and in case of multibody decays it may vary over phase space.

The mixing rate $R_{\rm{M}}$ defined as the time integral over the term III in Eq.~\ref{timeevolution} is independent of 
the strong phase ($ 
x^2 +y^2 = x'^2 +y'^2 $) and
can be measured in semileptonic $D^0$ decays.
\subsection{Event selection}
The $Q$ value of the $D^0$ production process $D^* \rightarrow D^0 \pi_{\rm{tag}} $ is about 6\,MeV  
which leads to a narrow peak in the difference of the $D^*$ and the $D^0$ mass ($\Delta m = m(D^0 \pi_{\rm{tag}}) - m(K\pi) $).
Selecting data in the peak region of the $\Delta m$ distribution suppresses backgrounds very effectively.
The $D^0$ mass reconstructed from identified kaon and pion tracks $m(K\pi)$ is required to be in the expected mass window and 
the $D^0$ momentum in the center-of-mass system (CMS) has to be larger than 2.5\,GeV in order to remove $D^0$ mesons from $B$ decays. 
The kaon and pion tracks are refit to originate from the same vertex and form a $D^0$ which is fit together with the slow  pion $\pi_{\rm{tag}}$
to a common vertex. This provides an event-wise measurement of 
the $D^0$ proper time $\tau$ and the error of the proper time $\sigma_{\tau}$. The typical average value is 240\,$\mu$m with a resolution of 100\,$\mu$m.

In the plane of $\Delta m$ and $m(K\pi)$, the {\slshape B\kern-0.1em{\smaller A}\kern-0.1emB\kern-0.1em{\smaller A\kern-0.2em R}} analysis
selects 1,129,000 RS and 64,000 WS candidates from a data sample of 384\,$\rm{fb^{-1}}$.

\section{Mixing measurements}
\subsection{\large{${\bf D^0 \rightarrow K \pi}$}}

The RS and WS event candidates in the plane of $\Delta m$  and $m(K\pi)$ contain different contributions of signal and backgrounds. Both are 
described by probability density functions (PDF). Their parameters are determined simultaneously for the RS and WS data sample 
in an unbinned maximum likelihood fit with four variables $ m (K\pi),\Delta m, \tau, \sigma_{\tau}$ per event. 
\begin{figure}[htb]
\begin{center}
\epsfig{file=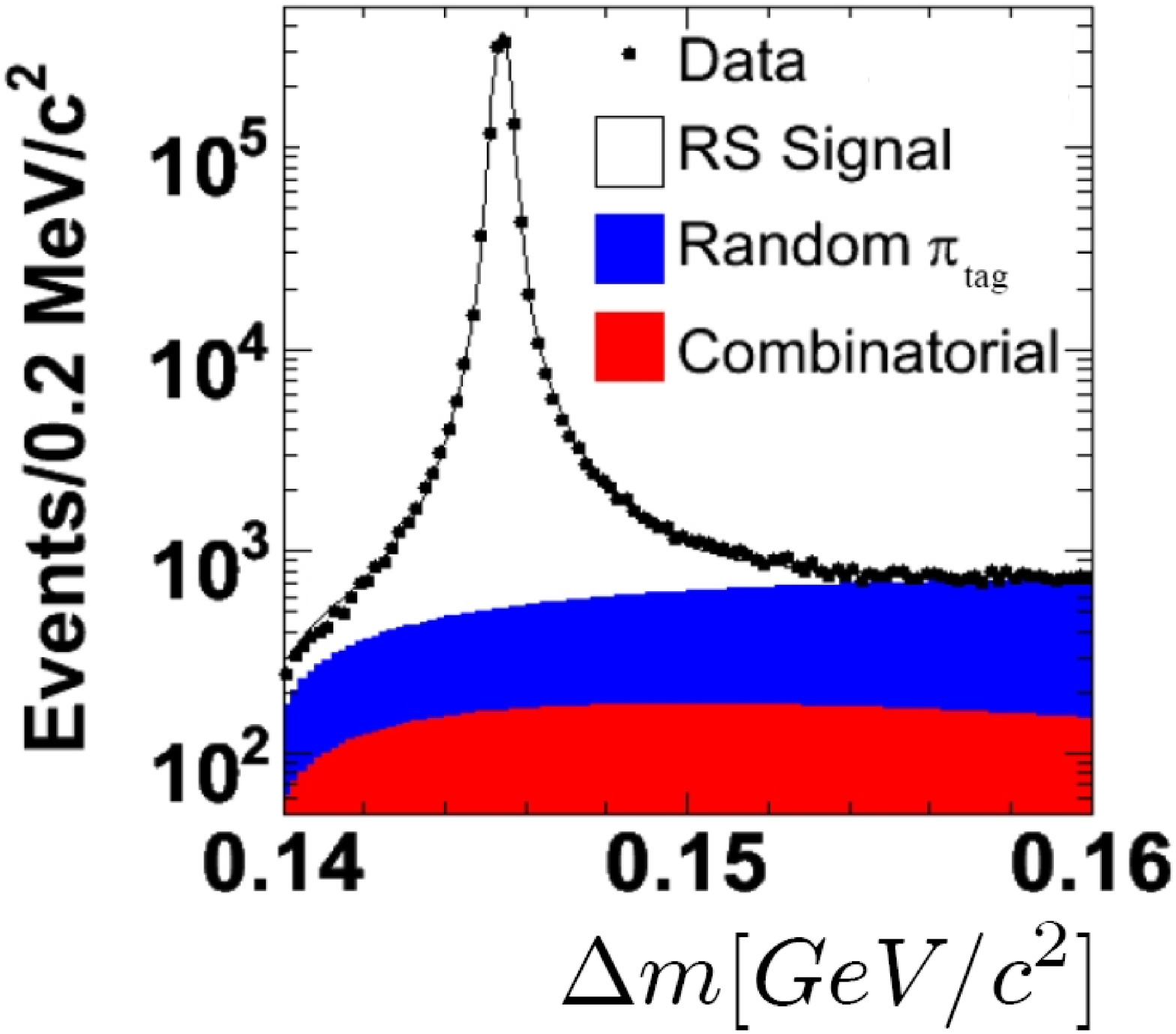,height=1.5in}
\hspace{0.8in}
\epsfig{file=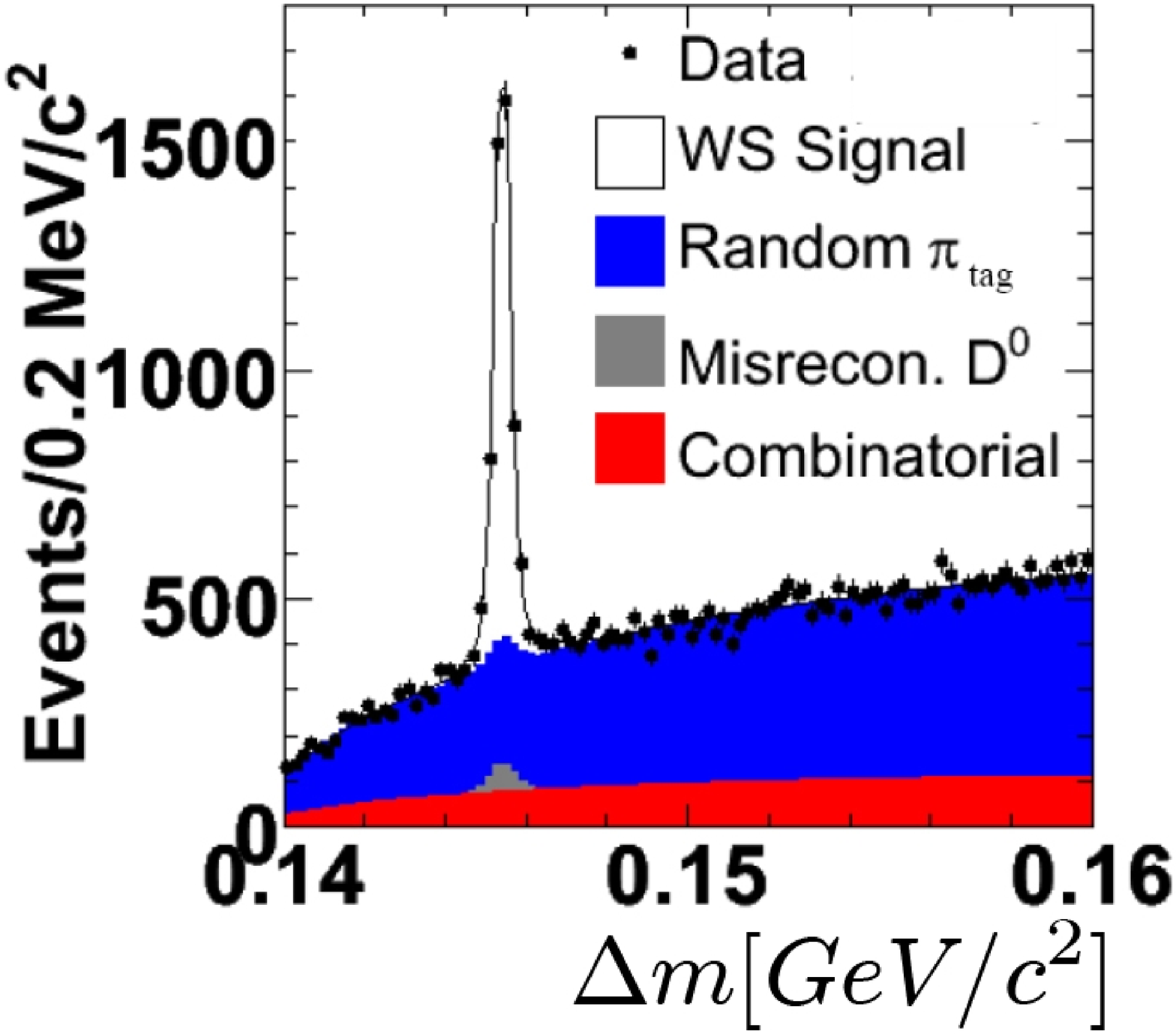,height=1.5in}
\\
\epsfig{file=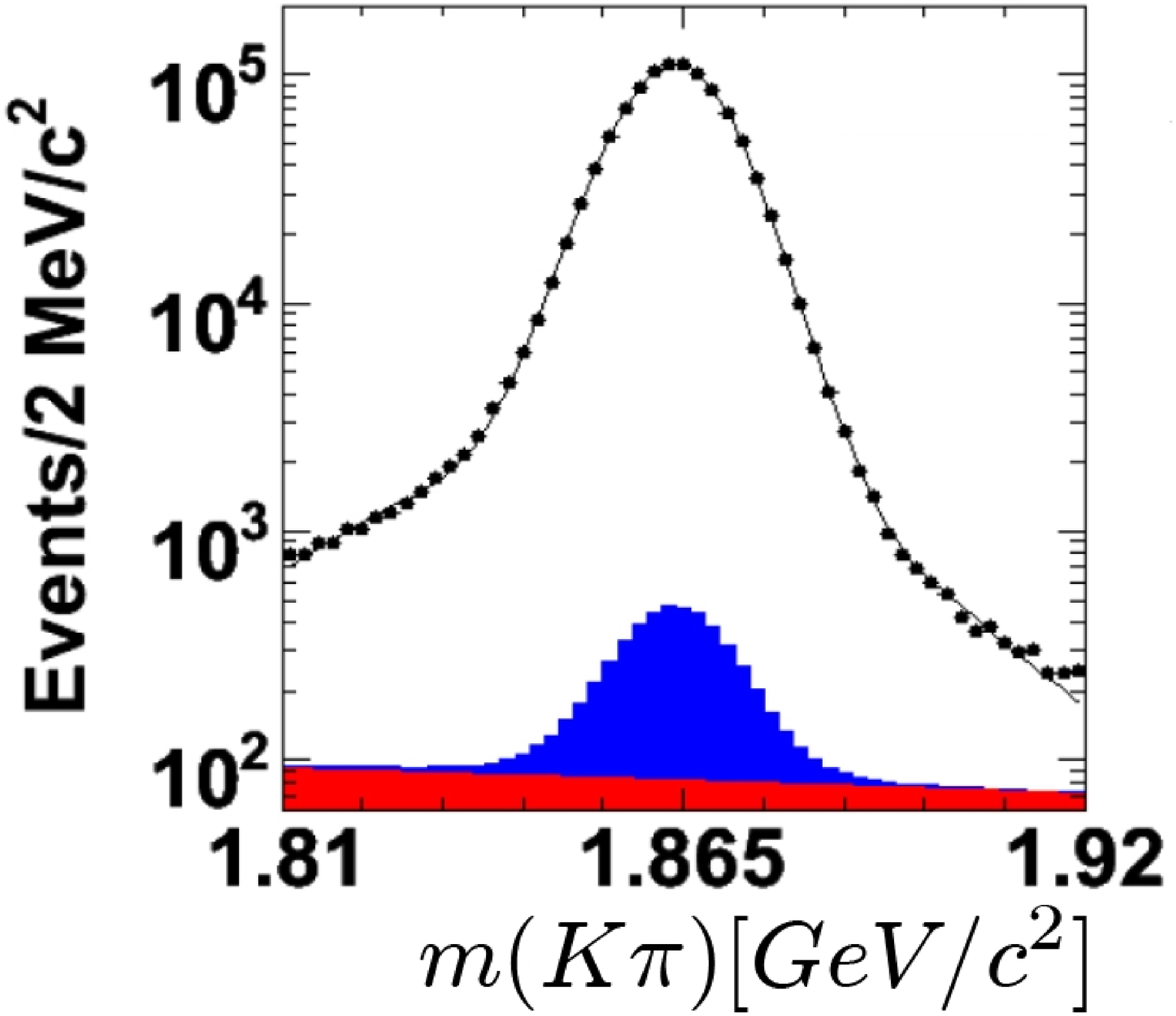,height=1.5in}
\hspace{0.8in}
\epsfig{file=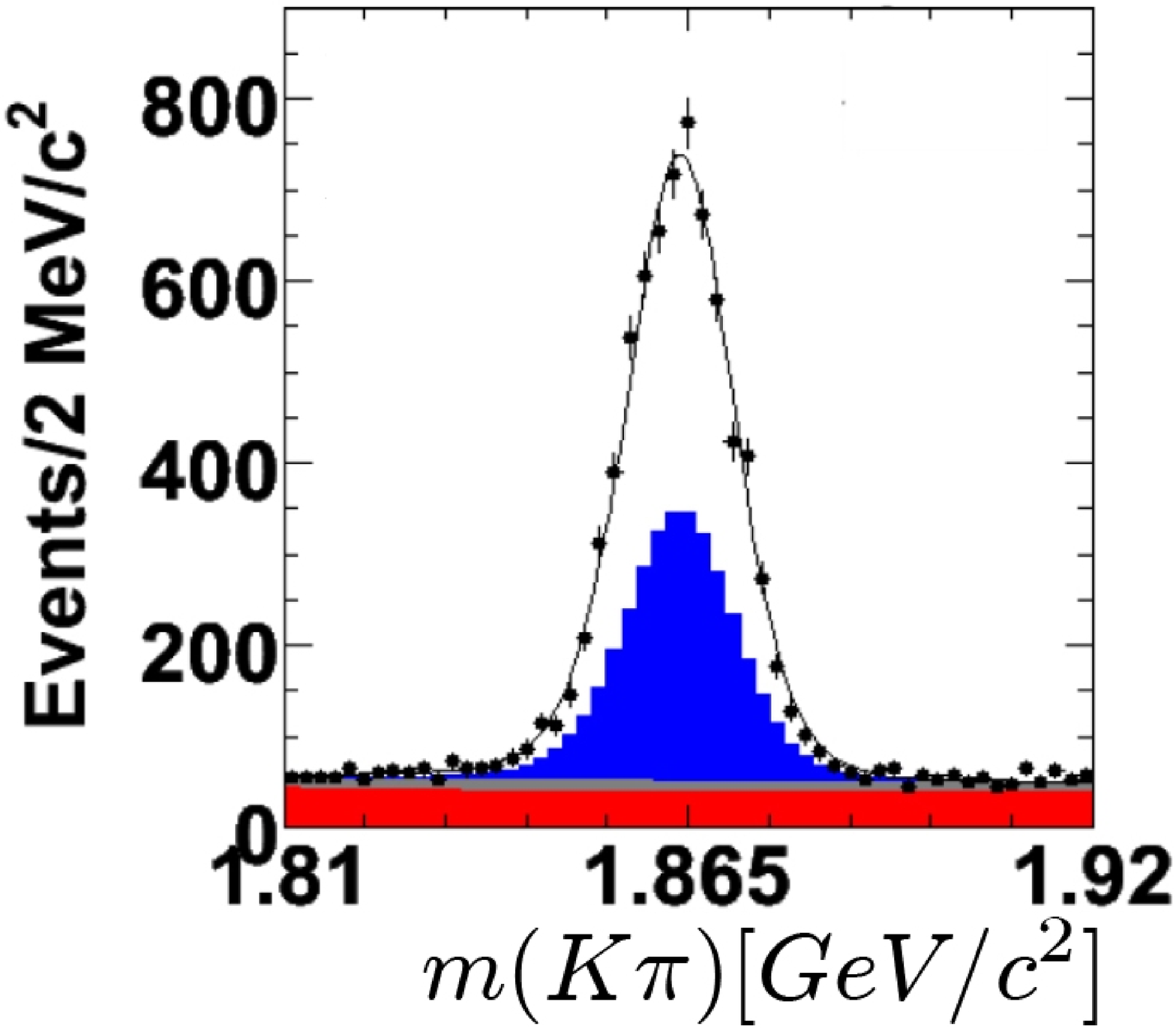,height=1.5in}
\caption{\label{Kpi_mdeltam}$m(K\pi)$ (bottom part) and $\Delta m(K\pi)$ (top part) for wrong-sign candidates (right) and right-sign candidates (left). The fitted PDF's are overlaid. The colored regions represent the different background components~\cite{d0discoverBABAR}.}
\end{center}
\end{figure}
Figure~\ref{Kpi_mdeltam} shows the projections in $m(K\pi)$ and in $\Delta m(K\pi)$ of the RS and WS 
candidates. The black points denote the data and 
the solid lines show the fitted PDF's. The white surface below the curve is the extracted signal.
The dominant background component (blue) originates from properly reconstructed $D^0$ mesons combined with a random slow pion $\pi_{\rm{tag}}$.

From the fitted signal yields, the WS branching fraction $R_{\rm{WS}}$ in the decay $D^0 \rightarrow K \pi$ is extracted. The {\slshape B\kern-0.1em{\smaller A}\kern-0.1emB\kern-0.1em{\smaller A\kern-0.2em R}} measurement~\cite{WSBranchingbabar} of 
 $R_{\rm{WS}} = (0.353 \pm 0.008 \pm 0.004) \%$  agrees 
well with the Belle result~\cite{WSBranchingbelle} of $R_{\rm{WS}} = (0.377 \pm 0.008 \pm 0.005) \%$, 
where the uncertainties are statistical and systematic, respectively. 

\begin{figure}[htb]
\begin{center}
\epsfig{file=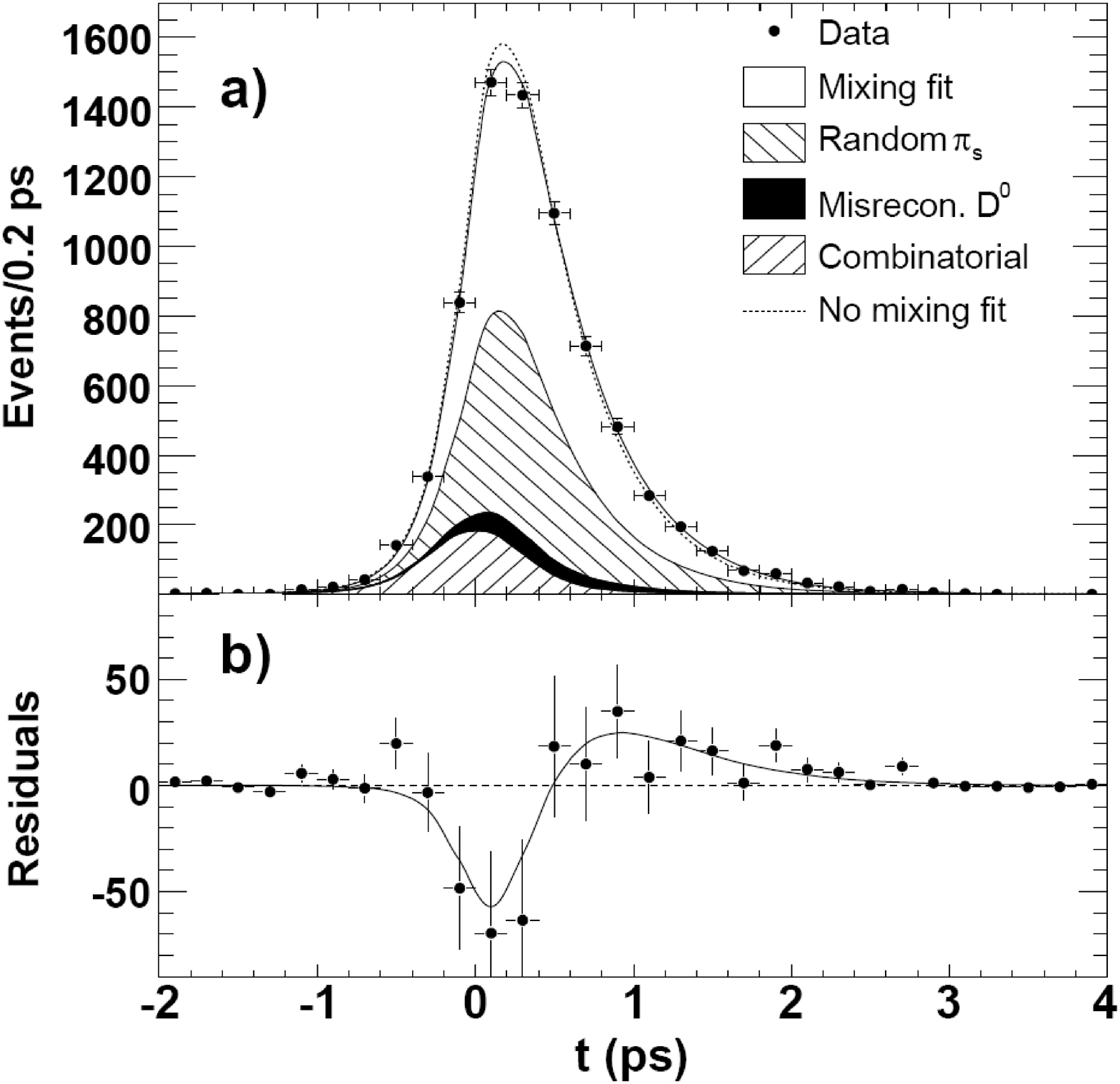,height=2.5in}
\caption{ \label{WSResidual} a) Projections of the proper time distribution of the WS candidates and of 
the fit results. The results of the
fit allowing (not allowing) mixing is overlaid as solid (dotted) curve. \hspace{0.1in}
b) The points represent the difference between the data and the no-mixing fit. The solid curve represents the difference between fits with and without mixing~\cite{d0discoverBABAR}.}
\end{center}
\end{figure}

\begin{figure}[htb]
\begin{center}
\epsfig{file=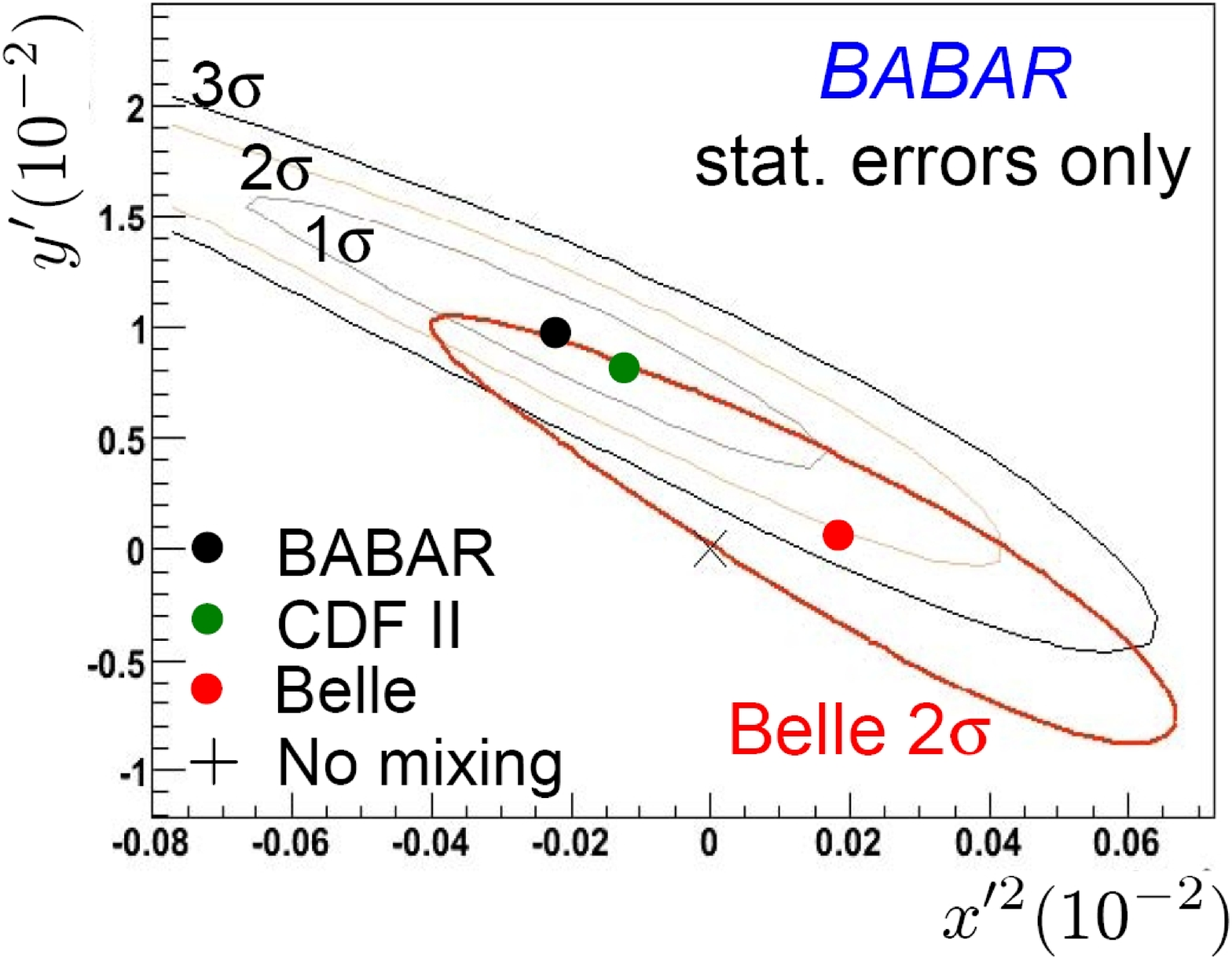,height=1.5in}
\hspace{1.0in}
\epsfig{file=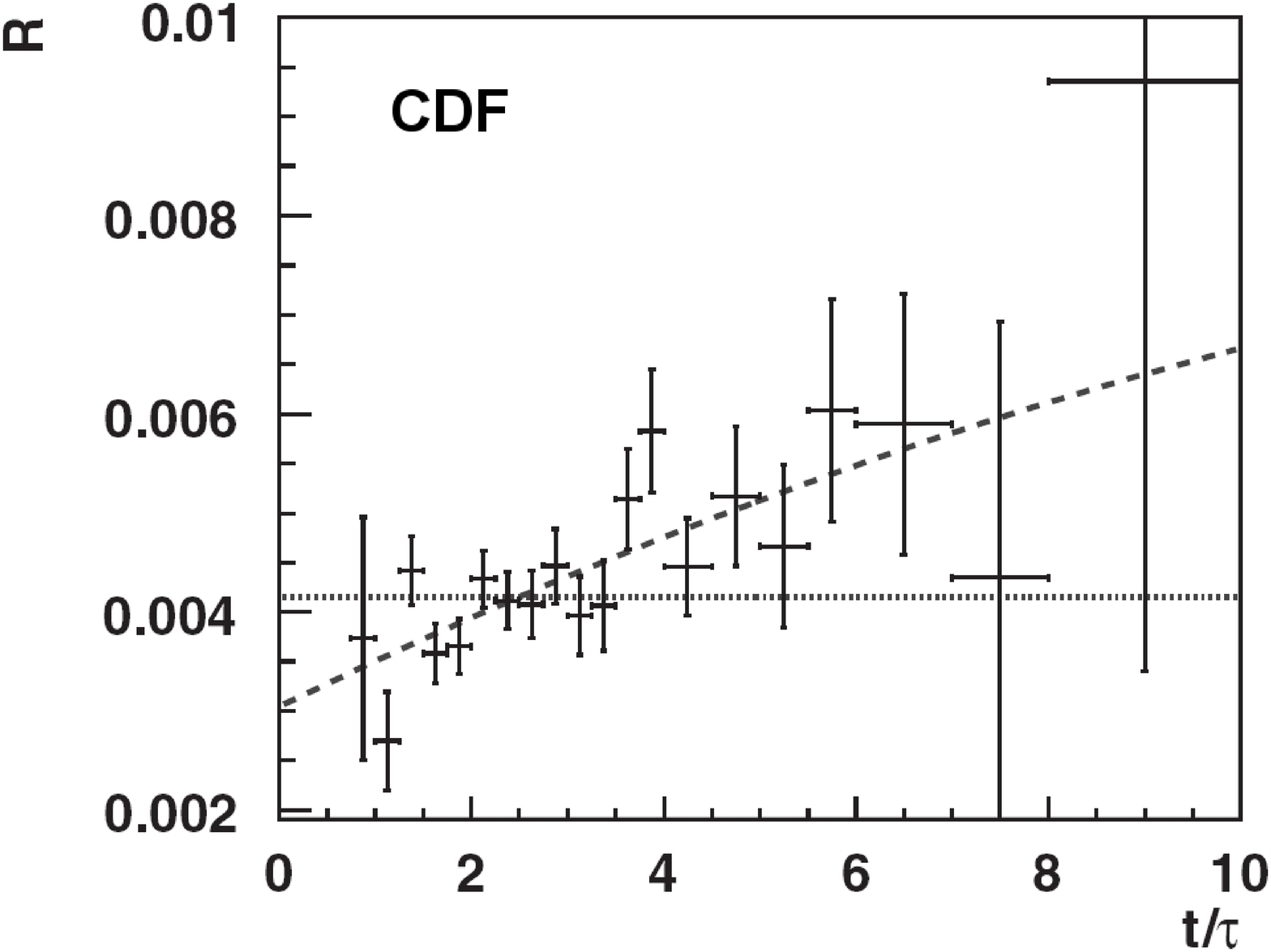,height=1.5in}
\caption{\label{y_x2_compare_Kpi} (Left) Comparison of the mixing parameters $x'^2$ and $y'$ obtained by {\slshape B\kern-0.1em{\smaller A}\kern-0.1emB\kern-0.1em{\smaller A\kern-0.2em R}} (black), Belle (red) and CDF II (green). The dark curves are the 1 $-$ 3 standard deviation likelihood contours from {\slshape B\kern-0.1em{\smaller A}\kern-0.1emB\kern-0.1em{\smaller A\kern-0.2em R}}, the red curve is the two standard deviation contour from Belle (stat. uncertainties only)~\cite{d0discoverBABAR,WSBranchingbelle,d0discoverCDF}. (Right) Ratio of WS and RS $D^0$ decays as function of normalized proper time as obtained by the CDF collaboration. The dashed (dotted) curve is the fit with the mixing (no-mixing) hypothesis~\cite{d0discoverCDF}.}
\end{center}
\end{figure}

The  WS signal contains mainly DCS events with a small fraction of mixing events. In order to separate the 
mixing signal, $D^0$ decay time information is used as described in Sec.~\ref{mixingparameter}. 
The $D^0$ lifetime and the lifetime resolution function is determined from the RS event sample. The WS event
sample is fit according to Eq.~\ref{timeevolution}. The result is shown as solid curve in Figure~\ref{WSResidual}.
The mixing parameters are measured to be
$R_{\rm{D}}=(0.303 \pm 0.016 \pm 0.01)\%$, $y'= (0.97 \pm 0.44 \pm 0.31)\%$ and 
$x'^2=(-0.022 \pm 0.03 \pm 0.021) \% $. The dotted curve shows the fit with the assumption of no-mixing.
The histogram in the lower part of Figure~\ref{WSResidual} displays the difference of the data and the no-mixing fit (dots),
while the curve is the difference between the mixing and the no-mixing fit model. 
The mixing model describes the deviations seen in the 
residuals.
 Thus, the deviations of the points from zero can be accounted for by the $D^0$ mixing.
The significance of the mixing signal is evaluated based on the change of the negative log-likelihood with respect to the maximum.
Figure~\ref{y_x2_compare_Kpi} (left part) shows the confidence-level contours calculated from the change in 
log-likelihood ($ - 2 \Delta \ln \mathcal{L}$)
in the two dimensions of $x'^2$ and $y'$ considering statistical uncertainties only. 
The likelihood maximum is indicated as black dot and is located in an unphysical region ($x'^2 < 0$).
The most likely physical allowed 
value ($x'^2=0$ and $y'= 6.4 \cdot 10^{-3}$) has a log-likelihood $ - 2 \Delta \ln \mathcal{L}$ of 0.7 units.  
The value of $ - 2 \Delta \ln \mathcal{L}$ for no-mixing is 23.9 units.
Including the systematic uncertainties, this corresponds to a significance of 3.9 standard deviations and thus, constitutes evidence for
mixing.

To search for CP violation, Eq.~\ref{timeevolution} is applied to the WS $D^0$ and $\bar{D}\,^0$
samples separately, fitting for the parameters $R_{\rm{D}}$, $x'^2$ and $y'$ for $D^0$ decays (+) and
$\bar{D}\,^0$ decays (-). Both sets of mixing parameters
($y'^{+}= (0.98 \pm 0.64 \pm 0.45)\% $,  
$x'^{+2}=(-0.024 \pm 0.043 \pm 0.03) \% $) 
and ($y'^{-}= (0.96 \pm 0.61 \pm 0.43)\% $,  
$x'^{-2}=(-0.020 \pm 0.041 \pm 0.029) \% $) are fully compatible with each other and differ by more than
three standard deviations from the no-mixing
hypothesis.    
The values $R_{\rm{D}} = \sqrt{R^+_{\rm{D}} R^-_{\rm{D}}} = (0.303 \pm 0.16 \pm 0.10) \% $ and 
$A_{\rm{D}} = (R^+_{\rm{D}}-R^-_{\rm{D}})/(R^+_{\rm{D}}+R^-_{\rm{D}}) = (-2.1 \pm 5.2 \pm 1.5)\%$ are calculated. $A_{\rm{D}}$ is fully compatible with zero. Both sets
of mixing parameters do not differ. Therefore,
no evidence for CP violation is observed.

At the end of the year 2007, the CDF II collaboration published evidence for mixing in $D^0 \rightarrow K \pi$ decays using an 
integrated luminosity 
of 1.5 $\rm{fb}^{-1}$ in $p \bar{p}$ collisions \cite{d0discoverCDF}. The measurement exploits the time dependence of the
 number of WS and RS $D^0$ decays.
In the range from 0.75 to 10 units of the $D^0$ decay time ($\Gamma t$), 
$(12.7 \pm 0.3) \cdot 10^3$ WS and $(3.044 \pm 0.002) \cdot 10^6$ RS signal events
are selected. A least-squares parabolic fit of Eq.~\ref{timeevolution} to $R = N_{\rm{WS}} / N_{\rm{RS}}$ 
in 20 bins of $\Gamma t$  as shown in the right part of Figure~\ref{y_x2_compare_Kpi} determines the mixing parameters 
$R_{\rm{D}}=(0.304 \pm 0.055)\%$, $y'= (0.85 \pm 0.76)\%$ and 
$x'^2=(-0.012 \pm 0.035) \% $. The no-mixing hypothesis is indicated as dotted line.  Despite the different production environment
 and analysis techniques, the agreement with the
{\slshape B\kern-0.1em{\smaller A}\kern-0.1emB\kern-0.1em{\smaller A\kern-0.2em R}}
mixing results is astonishing (see green dot in the left part of Figure~\ref{y_x2_compare_Kpi}). 
Bayesian probability contours in the $x'^2-y'$ plane are calculated. The data are inconsistent
with the no-mixing hypothesis with a probability equivalent to 3.8 standard deviations.    

The Belle measurement of mixing in $D^0 \rightarrow K \pi$  published in 2006 is based on an integrated luminosity of
400 $\rm{fb}^{-1}$~\cite{WSBranchingbelle}. Mixing parameters of $R_{\rm{D}}=(0.364 \pm 0.017)\%$, $y'= (0.06^{+0.40}_{-0.39})\%$ and 
$x'^2=(0.018^{+0.21}_{-0.23})\% $ are obtained. In Figure~\ref{y_x2_compare_Kpi} the results are compared to the 
{\slshape B\kern-0.1em{\smaller A}\kern-0.1emB\kern-0.1em{\smaller A\kern-0.2em R}} and CDF measurements.
Belle excludes the no-mixing hypothesis with a significance of 
two standard deviations considering statistical uncertainties only.   
\subsection{\label{strongPhase}Strong phase {\large ${\bf \delta_{K\pi}} $}}
The measurements of the mixing parameters $x$ and $y$ are only defined up to a strong phase $\delta_{K\pi}$
(amplitude ratio of the CF to the DCS decays in $D^0 \rightarrow K \pi$).
The CLEO collaboration recently published a measurement of $\delta_{K\pi}$
using quantum correlated $D^0–\bar{D}\,^0$
pairs\footnote{CLEO uses the inverse amplitude ratio for the definition of the strong phase $\delta_{K\pi}$.}
which were produced in 281 $\rm{pb}^{-1}$ of $e^+e^-$ collisions on the $\Psi$(3770) resonance~\cite{deltakpi}.
Two general decay classes are considered, single tags and double tags. For the single tags, one $D^0$ is reconstructed
independently of the other. This class provides uncorrelated decay information. 
In case of the double tags, both $D^0$ mesons are reconstructed, and they decay correlated. 
The final states can be hadronic or semileptonic.
Depending on the final state, different enhancement factors to the ratio of the
correlated and uncorrelated $D^0$ decay rates apply. The enhancement factors 
are functions of the mixing parameters and the strong
phase. Therefore, the measurement of the time integrated yields of the correlated and uncorrelated $D^0$ decays
allows to extract the mixing parameters and the strong phase by a fitting procedure. 
Using external branching fraction measurements in the fit, CLEO obtains $\cos(\delta_{K\pi}) = 1.03^{+0.31}_{-0.17} \pm 0.06$.
Including additional external measurements of the mixing parameters in the fit provides an alternate measurement 
of $\cos(\delta_{K\pi}) = 1.10 \pm 0.35 \pm 0.06 $ and allows for a determination
of $ x \cdot \sin(\delta_{K\pi}) = (4.4^{+2.7}_{-1.8} \pm 0.29) \cdot 10^{-3}$ and  $\delta_{K\pi} = (22\; ^{+11 + \; \; 9}_{-12 -11})^{\circ}$.

With these measurements, CLEO established a new technique of time-independent measurements of mixing parameters
and the first measurement of the strong phase.  
\subsection{\label{yCP}{\large${\bf y_{\rm{CP}}}$} from lifetime measurements}
The decay time $\tau$ of $D^0$ mesons (+) and  $\bar{D}\,^0$ mesons (-) decaying into final states of specific CP 
(such as $K^- K^+$  and $\pi^- \pi^+$)
can be considered to first order as exponential
with a small correction term that depends on the mixing parameters:
\begin{displaymath} 
\tau^{\pm} =  \frac {\tau^0 } {1+ |q/p| (y \, \cos \phi_{f}  \mp x \, \sin \phi_{f} ) }  \; \; ,
\end{displaymath}
where $\tau^0$ is the lifetime of the CF decay  $D^0 \rightarrow K \pi$.
The lifetimes can be combined into the quantities $y_{\rm{CP}}$ and $\Delta Y$:
\begin{displaymath}
y_{\rm{CP}} = \frac{\tau^0}{\tau} -1  = \frac{\tau (K^- \pi^+)}{\tau ( \pi^+ \pi^-) } -1 =  \frac{\tau (K^- \pi^+)}{\tau ( K^+ K^-) } -1  \; \; \; \; \; \; \; \; \; \; \; \; \; \; \; \Delta Y = \frac{\tau^0 A_{\tau}}{\tau} \;  , 
\end{displaymath}
with $A_{\tau}= (\tau^+ -\tau^-) /  (\tau^+ + \tau^-) $.
In the limit of CP conservation\footnote{The sign depends on the CP eigenvalue.}, $y_{\rm{CP}}= \pm y$ 
  and $\Delta Y = 0$.

Belle found significantly different decay time distributions for $D^0$ decays to the
CP-eigenstates $K^+ K^-$ and $\pi^+ \pi^-$ compared to the one to the CP-mixed state $K^-\pi^+$
and measured $y_{\rm{CP}} = (1.31\pm 0.32 \pm 0.25)\%$ with a significance of 3.2 standard deviations
including systematic uncertainties~\cite{d0discoverBELLE}. The dataset corresponds to 540~$\rm{fb}^{-1}$. 
The measured lifetime asymmetry parameter
$A_{\tau} = - (0.01 \pm 0.3 \pm 0.15) \cdot 10^{-3}$ reveals no evidence for CP violation.  

At the end of 2007, {\slshape B\kern-0.1em{\smaller A}\kern-0.1emB\kern-0.1em{\smaller A\kern-0.2em R}}
 published  a measurement of $y_{\rm{CP}}$ from the lifetime of the three $D^0$ decay modes $K^+K^-$, $\pi^+\pi^-$
 and $K \pi$ \cite{YCPBABAR} using 384 $\rm{fb}^{-1}$ of data. A value of $y_{\rm{CP}} = (1.24 \pm 0.39 \pm 0.13) \%$ is obtained,
which is evidence for mixing at the three standard deviation level. No indication for CP violation was found, as
indicated by the value $\Delta Y = (-0.26 \pm 0.36 \pm 0.08) \%$. Combining this result with a previous untagged  
{\slshape B\kern-0.1em{\smaller A}\kern-0.1emB\kern-0.1em{\smaller A\kern-0.2em R}} measurement \cite{YCPBABAROLD}
yields the combined measurement $y_{\rm{CP}} = (1.03 \pm 0.33 \pm 0.19) \%$.

The charm subgroup of the Heavy Flavour Averaging Group (HFAG) provides combined values for mixing parameters~\cite{HFAG}.
Besides 
{\slshape B\kern-0.1em{\smaller A}\kern-0.1emB\kern-0.1em{\smaller A\kern-0.2em R}} and Belle,
results from E791, FOCUS and CLEO contribute to the average values $y_{\rm{CP}} = (1.132 \pm 0.266) \%$ and 
$A_{\tau} = - (0.123 \pm 0.248 ) \% $.
The precision is dominated by the $B$ factory experiments. The $y_{\rm{CP}}$ measurement clearly indicates 
$D^0$ mixing at a lifetime
difference which is about $1\,\%$  and the no-mixing case is excluded at 4.5 standard deviations.
There is no indication for CP violation from the averaged asymmetry measurements.   
\subsection{\label{dalitz}Time-dependent Dalitz analysis}
Up to now, we considered only two body decays of the $D^0$. CLEO pioneered a method to 
measure $x$ and $y$ from a time-dependent Dalitz analysis of the resonant substructure in the decay $D^0 \rightarrow K^0_S \pi^+ \pi^-$~\cite{dalitzcleo}. This method allows to measure the sign of $x$.

\begin{figure}[htb]
\begin{center}
\epsfig{file=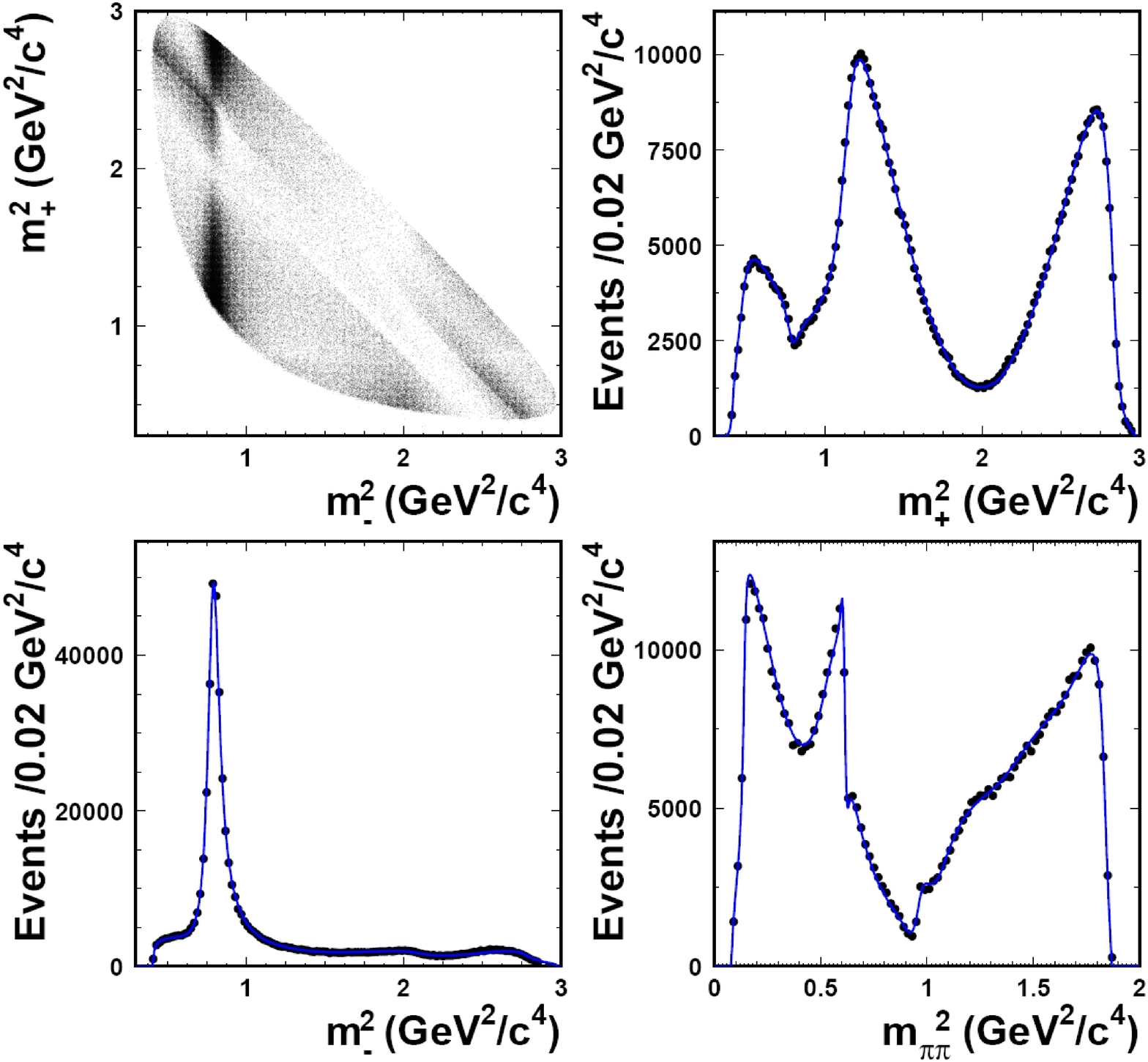,height=2.6in}
\caption{\label{DalitzPlot} Dalitz plot distribution and the projections in case of data (points with error bars) and the fit result (curve) as obtained by the Belle collaboration~\cite{dalitzbelle}.}
\end{center}
\end{figure}

Belle extends the analysis of the self-conjugate process $D^0 \rightarrow K^0_S \pi^+ \pi^-$ to a dataset of 540 $\rm{fb}^{-1}$~\cite{dalitzbelle}. 
The Dalitz plot is shown in the upper left distribution of Figure~\ref{DalitzPlot}. It
is described by a model containing 
contributions of 18 different quasi two body decays, which interfere. 
The time-dependent decay amplitudes of the $D^0$ and $\bar{D}\,^0$ are functions of the Dalitz variables ($m^2_+(K_S^0 \pi^+)$, $m^2_-(K_S^0 \pi^-)$) and the mixing parameters
$x$ and $y$. Therefore, $x$ and $y$ can be extracted from an unbinned maximum likelihood fit to the 
Dalitz variables and the measured $D^0$ decay time. The data points are shown in Figure~\ref{DalitzPlot} and the curves represent the fit results.
Assuming negligible CP violation, $x = ( 0.80 \pm 0.29 ^{+0.09 +0.10}_{-0.07-0.14} ) \%$ and 
$y =  ( 0.33 \pm 0.24 ^{+0.08+0.06}_{-0.12-0.08} ) \%$ are obtained, where the uncertainties are statistical,
experimental systematic and decay-model systematic. 
The no-mixing case is excluded at 2.2 standard
deviations. 

Allowing for CP violation, the fit of the additional parameters $|q/p|$ and $\phi$ indicates
no evidence for CP violation in mixing or interference between mixed and unmixed amplitudes.
Since the fit parameters are consistent for both the $D^0$ and $\bar{D}\,^0$ sample, there is also
no evidence for direct CP violation.  

{\slshape B\kern-0.1em{\smaller A}\kern-0.1emB\kern-0.1em{\smaller A\kern-0.2em R}}
finds evidence for $D^0$ mixing using a time-dependent amplitude analysis  of the
decay  $D^0 \rightarrow K^+ \pi^- \pi^0$  in a data sample of 384 $\rm{fb}^{-1}$~\cite{dalitzbabar}.
The decay contains WS and RS events (see Sec.~\ref{flavortagging}). The signal and background yields
are extracted from a binned extended maximum likelihood fit to the  $\Delta m$ and  $m(K\pi\pi^0)$  distributions.
The time-dependent relative WS decay rate is a function of the Dalitz variables ($m^2_{K^+ \pi^-}$, $m^2_{K^+ \pi^0}$) 
and the mixing parameters in the form $x'_{K \pi \pi^0} = x \cos \delta_{K \pi \pi^0} + y \sin \delta_{K \pi \pi^0} $ 
and  $y'_{K \pi \pi^0} = y \cos \delta_{K \pi \pi^0} - x \sin \delta_{ K \pi \pi^0}$,
where $\delta_{K \pi \pi^0}$ is the strong phase between the DCS and the CF amplitude of $D^0 \rightarrow \rho^- K^+$. The phase $\delta_{K \pi \pi^0} $ is different
from $\delta_{K \pi} $ and has to be measured elsewhere. From a time-dependent
fit to the WS data, $x'_{K \pi \pi^0} =(2.61^{+0.57}_{-0.68}\pm 0.39) \% $  and   $y'_{K \pi \pi^0} =(-0.06^{+0.55}_{-0.64}\pm 0.34) $ are
derived with a correlation of $-0.75$. The significance is equivalent to 3.2 standard deviations.

\section{Combined results}
The HFAG
determined world average values of the mixing parameters $x$ and $y$
 in a global fit which takes  into account all the
relevant data from the various 
experiments~\cite{HFAG}. Most of the more
recent results have been presented in this talk. They are dominated by the $B$ factory measurements
with significant contributions from CLEO-c and Tevatron. 
The no-mixing case is excluded at about seven standard deviations ($ x = (0.91 \pm 0.26) \% $ and $y = (0.73 \pm 0.18) \% $). The
mass difference differs from zero by 3 standard deviations, the lifetime difference deviates from zero by about
4.1 standard deviations.
The fit also determines $R_{\rm{D}}= (0.3342  \pm 0.0083) \%  $, $\delta_{K\pi} = (21.6^{+11.6}_{-12.6})^{\circ}$ and $\delta_{K\pi \pi^{0}} = (30.8^{+25.0}_{-25.8})^{\circ}$.
The measurement of CLEO of the strong hadronic phase $\delta_{K\pi}$ discussed in \ref{strongPhase} was 
not considered because external 
measurements of $R_{\rm{D}}$ and $R_{\rm{M}}$ entered in the fitting procedure.

Another set of parameters was determined allowing for CP violation. The values obtained for the CP sensitive parameters $A_{\rm{D}}=-2.2 \pm 2.5$, $|q/p|= 0.86^{+0.18}_{-0.15}$ and $\phi = (-9.6^{+8.3}_{-9.5})^{\circ}$ indicate no 
evidence for CP violation within the current sensitivity of the experiments.  

The $y$ measurement from the $D^0$ lifetime (see Sec.~\ref{yCP}) yields $y > 0$ .
Therefore, the $|D_1\rangle$ as CP-even state lives shorter than the CP-odd state $|D_2\rangle$.
The sign of $x$ is measured in the Dalitz analysis (see Sec.~\ref{dalitz}) as $x > 0$. Therefore, the
CP-even state $|D_1\rangle$ is heavier than the CP-odd state $|D_2\rangle$ .

Preliminary Monte Carlo studies of the LHCb collaboration indicate that the statistical uncertainties of the mixing parameters $x, y$ and $y_{\rm{CP}}$ measured in a time-dependent
WS analysis of $D^0 \rightarrow  K\pi$ decays and a lifetime analysis of $D^0$ decays to $K^-K^+$ and $K^-\pi^+$ in 10 $\rm{fb}^{-1}$ of data may decrease by a factor five~\cite{lhcb}.   

In summary, three experiments found evidence for $D^0$ mixing measuring lifetime and 
mass differences at the level of 1 \%. The combined results of all experiments exclude
the no-mixing case at seven standard deviations. The measurements are compatible with
the SM expectations. It seems likely that the $D^0$ mixing is dominated by
long-distance processes, which are difficult to calculate. Therefore, identifying NP
contributions from mixing alone is not easily possible. Neither a single experiment nor
the combination of all results provide any hint for CP violation in $D^0$ mixing.

\end{document}